\documentclass[twocolumn]{jpsj2}

\def\runauthor{
Akira \textsc{Oguri} 
and A. C. \textsc{Hewson}
}

\title{
NRG approach to the transport through a finite Hubbard chain 
connected to reservoirs
}

\author{
Akira \textsc{Oguri} 
and A. C. \textsc{Hewson}$^1$
}%

\inst{%
Department of Material Science, 
Osaka City University, Sumiyoshi-ku, Osaka 558-8585, Japan \\
$^1$Department of Mathematics, Imperial College, 
180 Queen's Gate, London SW7 2BZ, UK
}%

\recdate{December 9, 2004}

\abst{
We study the low-energy properties of a Hubbard chain 
of finite size $N_C$ connected to two noninteracting leads using 
the numerical renormalization group (NRG) method.
The results obtained for $N_C=3$ and $4$  show that  
the low-lying eigenstates have one-to-one correspondence 
with the free quasi-particle excitations of a local Fermi liquid.
It enables us to determine the transport coefficients 
from the fixed-point Hamiltonian.
At half-filling, the conductance for even $N_C$ 
decreases exponentially with increasing $U$ 
showing a tendency towards the development 
of a Mott-Hubbard gap. 
In contrast, for odd $N_C$, 
the Fermi-liquid nature of the low-energy states assures 
 perfect transmission through the Kondo resonance.  
Our formulation to deduce the conductance from 
the fixed-point energy levels can be applied  
to various types of interacting systems.

}

\kword{
Kondo effect, Hubbard model, reservoir,
Fermi liquid, quantum dot, 
numerical renormalization group
}

\begin{document}
\sloppy

\maketitle

\section{Introduction}
\label{sec:introduction}

Electron transport through finite systems, such as 
quantum dots, quantum wires, and atomic chains of nanoscale,   
is a subject of much current interest. In these systems, 
a number of phenomena have been predicted theoretically 
and some have already been successfully observed. 
The Kondo effect in quantum dots is one such example.
\cite{GlatzmanRaikh,NgLee,Goldharber,Cronenwett}
Furthermore, recent experimental developments  make 
it possible to examine the interplay of various effects 
which have previously only  been studied 
in different fields of physics. 
For instance, in quantum dots, the 
interplay of the Aharanov-Bohm, Fano, Josephson, and Kondo 
effects under equilibrium and nonequilibrium situations 
have been studying intensively.\cite{Hofstetter,Kobayashi,SC}
The Luttinger-liquid behavior in quantum wires  
has also been an active field of investigation,\cite{KaneFisher,Tarucha}
and numerical developments have been reported 
recently for spinless fermions 
on a lattice.\cite{Meden,Molina}

\begin{figure}[bt]
\begin{center}
\setlength{\unitlength}{0.7mm}

\begin{picture}(130,20)(-5,0)
\hspace{-0.8cm}
\thicklines

\put(15,4){\line(1,0){25}}
\put(15,16){\line(1,0){25}}
\put(40,4){\line(0,1){12}}

\put(91,4){\line(1,0){25}}
\put(91,16){\line(1,0){25}}
\put(91,4){\line(0,1){12}}

\multiput(41.0,10)(2,0){4}{\line(1,0){1}}
\multiput(82.9,10)(2,0){4}{\line(1,0){1}}

\put(50,10){\circle*{4}} 
\put(60,10){\circle*{4}} 
\put(70,10){\circle*{4}} 
\put(80,10){\circle*{4}} 

\put(52,10){\line(1,0){6}}
\put(62,10){\line(1,0){6}}
\put(72,10){\line(1,0){6}}

\put(54,15){\makebox(0,0)[bl]{\Large $t$}}
\put(64,15){\makebox(0,0)[bl]{\Large $t$}}
\put(74,15){\makebox(0,0)[bl]{\Large $t$}}
\put(42.5,15){\makebox(0,0)[bl]
{\Large $v_L^{\phantom{\dagger}}$}}
\put(82,15){\makebox(0,0)[bl]
{\Large $v_R^{\phantom{\dagger}}$}}

\put(49,-0.5){\makebox(0,0)[bl]{\Large $1$}}
\put(59,-0.5){\makebox(0,0)[bl]{\Large $2$}}
\put(66.5,-0.5){\makebox(0,0)[bl]{\Large $\cdots$}}
\put(77.5,-0.5){\makebox(0,0)[bl]{\Large $N_C$}}

\end{picture}
\caption{Schematic picture of system.}
\label{fig:model}
\end{center}
\end{figure}
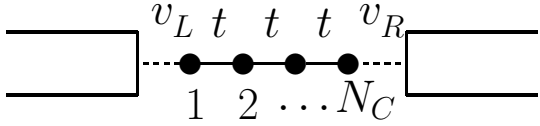

We have previously considered   
the transport properties of a finite Hubbard chain 
 of size $N_C$ (=$1$, $2$, $3$, \ldots ) connected 
to two noninteracting leads, which is illustrated in Fig.\ \ref{fig:model}, 
as a model for a series of quantum dots and materials on a nanometer scale. 
We have discussed an even-odd oscillatory behavior 
of the dc conductance at half-filling based on 
a  perturbation expansion in powers of 
the Coulomb repulsion $U$.\cite{ao8,ao9,ao10} 
For even $N_C$, the conductance decreases 
with increasing $U$ showing 
a tendency towards the development of 
a Mott-Hubbard insulator gap.\cite{ao8,ao9,ao10} 
The conductance deduced from the order $U^2$ self-energy 
was qualitatively correct.  
However, a more accurate treatment is needed to deal quantitatively 
with the large $U$ regime, which is  one of the aims of this paper.

In contrast, for odd $N_C$, the transmission probability through 
the Hubbard chain reaches the unitary-limit value at $T=0$ 
when the system has the electron-hole and inversion symmetries.
Physically there is a Kondo resonance situated 
at the Fermi level which enables the perfect transmission
take place for any value of $U$.
The proof was given by taking all contributions in powers of  $U$ 
formally into account for the $N_C \times N_C$ matrix self-energy 
$\mbox{\boldmath $\Sigma$} (\omega)$. The assumption we have made 
is; $\mbox{Re}\, \mbox{\boldmath $\Sigma$} (0)$ is {\em not singular} 
and  $\,\mbox{Im}\, \mbox{\boldmath $\Sigma$} (0) =0$ at $T=0$.\cite{ao9}  
The local Fermi-liquid state satisfies this assumption,\cite{ao5} 
and historically the same assumption has been 
made by Langer and Ambegaokar in the derivation of the Friedel sum rule 
for interacting electrons.\cite{LangerAmbegaokar}  
In the case of the single Anderson impurity corresponding to $N_C=1$, 
the perturbation theory in powers of $U$ describes the 
low-energy Fermi-liquid behavior,\cite{YamadaYosida} 
and is consistent with the exact 
Bethe ansatz\cite{KawakamiOkiji,WiegmanTsvelick,ZlaticHorvatic} 
and NRG\cite{Wilson,KWW,KWW2} results.
The perturbation expansion in $U$ works  
because the contributions from the low-energy processes, 
in which electrons  hop into the reservoirs and 
 away from the impurity, 
are included in the noninteracting Green's functions 
via the hybridization energy scale  $\Gamma_{L}$ and $\Gamma_R$,
where  $\Gamma_{L/R} = \pi \rho\, v_{L/R}^2$.
The perturbation expansion is also 
convergent for $N_C>1$.\cite{ao8,ao9,ao10}
It would be of interest to have these results confirmed
by a non-perturbative technique, which is the second aim of the work
presented in this paper.   

To tackle both these problems we apply the non-perturbative 
NRG approach to the low energy physics of the Hubbard chain,
connected to non-interacting leads. In doing so we  go beyond
the earlier low order perturbational results for the even site chains,
and we also derive 
 a Fermi liquid picture for the odd site chains,
without  making the  assumptions implicit in  the perturbation theory.  
The NRG method has been applied successfully 
to the quantum dots for $N_C=1$ and $2$.\cite{Izumida1-3,IzumidaPRB}
In the present work we have applied the NRG method for the Hubbard chain 
with size  $N_C=3$ and $4$, which seem to capture the essence 
of the even- and odd-size chains, respectively.
The results show that 
the low-lying energy states have a one-to-one correspondence with 
the quasi-particles excitations of the local Fermi liquid. 
It assures the validity of the Fermi-liquid description at low energies.
Specifically, for odd $N_C$, 
a number of the noninteracting sites are needed 
to be taken into account to reach the fixed point 
that describes the physics below the Kondo temperature $T_K$. 
The fixed point has much information about the low-temperature properties, 
and one can deduce the parameters such as $T_K$ 
and  Wilson ratio $R$ from the flow 
of the eigenvalues.\cite{Hewson_jphys,HewsonOguriMayer}
In the present paper, we provide a formulation to determine the conductance 
for even $N_C$ at $T=0$. 
 For large $U$, the NRG method improves the perturbation results,\cite{ao9}
and the conductance determined from the fixed-point energy levels
decreases exponentially with increasing $U$.

In \S \ref{sec:MODEL},
we deduce expressions of the ground-state properties 
 in terms of the Green's function.   
In \S \ref{sec:flow_vs_g},
 we describe the formulation to deduce  
the conductance from the fixed-point Hamiltonian.
In \S \ref{sec:NRG_results}, we show the NRG results.
In \S \ref{sec:remarks}, discussion and summary are given.

\section{Model and Formulation}
\label{sec:MODEL}

We consider a Hubbard chain of a finite size $N_C$ situated 
at the center, which is attached to two non-interacting leads 
at the left ($L$) and the right ($R$).
The complete Hamiltonian is given by 
\begin{align}
{\cal H}   & \, =\,    {\cal H}_{d} \, 
+ \, {\cal H}_{U} 
+ \,  {\cal H}_{\rm mix} \, + \, 
{\cal H}_{\rm lead} \;,
\label{eq:H}
\\             
 {\cal H}_{d}  & \, = \,  
 -t\sum_{i=1}^{N_C-1} 
 \sum_{\sigma} \left(  
 d^{\dagger}_{i\sigma}d^{\phantom{\dagger}}_{i+1\sigma} 
 \, +\, 
 d^{\dagger}_{i+1\sigma} 
 d^{\phantom{\dagger}}_{i\sigma}
 \right) 
 \nonumber \\
&  \ \quad  + \ 
 \sum_{i=1}^{N_C} 
    \left( \epsilon_d \,+\, \frac{U}{2} \right)  n_{d i} \;, 
    \\ 
 {\cal H}_{U}  & \, = \,  \frac{U}{2} 
 \sum_{i=1}^{N_C} \Bigl(\, n_{d i} -1\,\Bigr)^2 \;,
\label{eq:Hd}
\\
 {\cal H}_{\rm mix} \,  & \, =\,
v_{L}^{\phantom 0} 
 \sum_{\sigma}   
 \left(\,  
  d^{\dagger}_{1,\sigma} \psi^{\phantom{\dagger}}_{L \sigma}
\,+\,    
\psi^{\dagger}_{L \sigma}   d^{\phantom{\dagger}}_{1,\sigma} 
\,\right)
  \nonumber \\
& \ \ + v_{R}^{\phantom 0} 
 \sum_{\sigma}   
\left(\, 
\psi^{\dagger}_{R\sigma} d^{\phantom{\dagger}}_{N_C, \sigma} 
      \,+\,   
d^{\dagger}_{N_C, \sigma} \psi^{\phantom{\dagger}}_{R\sigma}
      \,\right)   ,
\label{eq:Hmix}
\\
 {\cal H}_{\rm lead}  & \, = \,  
\sum_{\nu=L,R} 
 \sum_{k\sigma} 
  \epsilon_{k \nu}^{\phantom{0}}\,
         c^{\dagger}_{k \nu \sigma} 
         c^{\phantom{\dagger}}_{k \nu \sigma}
\,,
\label{eq:H_lead}
\end{align}
where $d_{i\sigma}$ annihilates an electron with spin 
$\sigma$ at site $i$, 
and $n_{d i} = \sum_{\sigma}d^{\dagger}_{i\sigma} 
d^{\phantom{\dagger}}_{i\sigma}$. 
In the lead at $\nu$ ($= L,\, R$), the operator 
$c_{k \nu \sigma}^{\dagger}$ creates an electron 
with energy $\epsilon_{k\nu}$ corresponding to 
an one-particle state $\phi_{k\nu} (r)$. 
The hopping matrix elements
 $v_L^{\phantom{\dagger}}$ 
and $v_R^{\phantom{\dagger}}$ connect 
the chain and leads.
At the interfaces, 
a linear combination of the conduction electrons
 $\psi_{\nu \sigma}^{\phantom{\dagger}} 
= \sum_k c_{k \nu \sigma}^{\phantom{\dagger}} 
\, \phi_{k\nu} (r_{\nu}) $ mixed with the electrons at  $i=1$ or $N_C$ 
(as illustrated in Fig.\ \ref{fig:model}), 
where $r_{\nu}$ denotes the position at the interface in the lead side.

For this system, the Green's function is defined by
\begin{equation} 
G_{jj'}(i\omega_n) 
\, =\,  
-    \int_0^{\beta} \! \text{d}\tau \,
   \left \langle \, T_{\tau} \,  
   d^{\phantom{\dagger}}_{j \sigma} (\tau) 
   \, d^{\dagger}_{j' \sigma} (0) 
   \,  \right \rangle  \, e^{\text{i}\, \omega_n \tau} \;,
\label{eq:G_Matsubara}
\end{equation} 
where $\beta= 1/T$,  
$d_{j \sigma}(\tau) = e^{\tau  {\cal H}} d_{j \sigma} e^{- \tau  {\cal H}}$, 
and $\langle \cdots \rangle$ denotes the thermal average 
$\mbox{Tr} \left[ \, e^{-\beta  {\cal H} }\, {\cdots}
\,\right]/\mbox{Tr} \, e^{-\beta  {\cal H} }$.
We use units $\hbar=1$. 
The corresponding retarded function,  
 $G_{jj'}^{+}(\omega) =  G_{jj'}(\omega+ \text{i}\, 0^+)$, 
 is obtained via the analytic continuation. 
Since the interaction $U$ is finite only 
for the electrons in the chain at the center,
the Dyson equation is written in the form
\begin{equation} 
  G_{ij}(z)  \,  = \,  G^0_{ij}(z) 
   \, +  \sum_{i'j' =1}^{N_C}\,G^0_{ii'}(z)\,  \Sigma_{i'j'}(z)
   \, G_{j'j}(z) \;.
  \label{eq:Dyson}
\end{equation} 
Here $G^0_{ij}(z)$ is the unperturbed Green's function 
corresponding to ${\cal H}_0 \equiv {\cal H}_d 
+ {\cal H}_{\rm lead} + {\cal H}_{\rm mix}$, 
and $\Sigma_{ij}(z)$ is the self-energy correction 
due to ${\cal H}_U$. 
Note that 
$G_{ij}(z) =  G_{ji}(z)$, 
because of the time reversal symmetry of ${\cal H}$. 
The Dyson equation  can be rewritten 
 using $N_C \times N_C$ matrices 
  $\mbox{\boldmath $G$}(z) = \{G_{ij}(z)\}$  
and $\mbox{\boldmath $\Sigma$}(z) = \{\Sigma_{ij}(z)\}$ as  
\begin{equation}
\left\{\mbox{\boldmath $G$}(z)\right\}^{-1}  
  \, = \,   
z \, \mbox{\boldmath $1$} 
-  \mbox{\boldmath $H$}_C^0  
- \mbox{\boldmath $V$}_{\rm mix}(z)   
- \mbox{\boldmath $\Sigma$}(z)   \;,
\label{eq:G_matrix}
\end{equation}
where
\begin{align}
 \mbox{\boldmath $H$}_C^0  &\, = \,    
\left [ \ 
%% \matrix {
\begin{matrix}
  0       & -t      &  &    \mbox{\Large $0$}           \cr
 -t       & 0 & \,  \ddots       &               \cr
          & \, \ddots              & \ddots  & -t  \cr 
\mbox{\Large $0$} &               &  -t       & 0 \cr
\end{matrix}
%%          }
          \  \right ]  
\;, 
\\
\mbox{\boldmath $V$}_{\rm mix}(z) &\,  =  \,
 \left [ \, 
%% \matrix { 
\begin{matrix}
 v_L^{2} \mbox{\sl g}_L^{\phantom{\dagger}}(z) & 0 &  & \mbox{\Large $0$} \cr
 0           &    0     & \ \ddots   &      \cr
      &  \ddots \ & \ddots   &  0  \cr 
\mbox{\Large $0$} &  & 0 & v_R^{2} \mbox{\sl g}_R^{\phantom{\dagger}}(z) \cr
\end{matrix}
%%}
 \, \right ]  
 \;,
\label{eq:V_mix}
\end{align}
and   $\mbox{\sl g}_{\nu}^{+}(\omega) \equiv \sum_k
{
\left| \phi_{\nu \sigma}^{\phantom{\dagger}} (r_{\nu}) \right|^2
/ 
(\omega - \epsilon_{k\nu}+ i0^+) 
}$ is the Green's function at interface of the isolated lead.
In the present study, 
we assume that the density of states is a constant, 
$\mbox{\sl g}_{\nu}^{+}(\omega) = -\text{i}\pi \rho_{\nu}$, for small $\omega$.
Then the energy scale of the level-broadening 
becomes $\Gamma_{\nu}= \pi\rho_{\nu} v_{\nu}^2$, 
and it determines the two non-zero elements of
 $\mbox{\boldmath $V$}_{\rm mix}^+(\omega)$.

\subsection{Ground-state properties}
\label{subsec:GS}

If the ground state has a 
property $\mbox{Im}\, \mbox{\boldmath $\Sigma$}^{+} (0) =0$  at $T=0$,  
the damping of the excitations at the Fermi level vanishes.
Then the effective Hamiltonian defined by
\begin{equation}
\mbox{\boldmath $H$}_C^{\rm eff}  
\, \equiv\, 
   \mbox{\boldmath $H$}_C^0  
            + \mbox{Re}\, \mbox{\boldmath $\Sigma$}^+(0)
\label{eq:K} 
\end{equation}
plays a central role on the ground-state properties.
It determines the renormalized hopping matrix elements 
$\mbox{\boldmath $H$}_C^{\rm eff} =\{ -\widetilde{t}_{ij}\, \}$, 
and also the value of the Green's function 
at the Fermi level 
$\left\{\mbox{\boldmath $G$}^+(0)\right\}^{-1}  = 
\mbox{\boldmath $K$}(0)  -  \mbox{\boldmath $V$}_{\rm mix}^+(0)$, where 
 \begin{equation}
\mbox{\boldmath $K$}(\omega) \,\equiv\,  
\omega \,\mbox{\boldmath $1$} 
-    
\mbox{\boldmath $H$}_C^{\rm eff}  
\;. 
\label{eq:K_omega} 
\end{equation}
The determinant of 
the matrix $\left\{\mbox{\boldmath $G$}^+(0)\right\}^{-1}$ is 
related to the scattering matrix, 
and can be rewritten in the following form 
by expanding the first and $N_C$-th columns, 
\begin{align}
\det \left\{\mbox{\boldmath $G$}^+(0)\right\}^{-1} 
 & \,=\,         \left[ \, 
        -  \Gamma_L \,\Gamma_R \, \det \mbox{\boldmath $K$}_{11}^{N_CN_C}(0) 
        + \det \mbox{\boldmath $K$}(0) 
       \, \right] \, 
      \nonumber \\
&
\!\!\!\!\!\!
+ \, \text{i}\, 
        \Bigl[ \,\Gamma_L \, \det \mbox{\boldmath $K$}_{11}(0) + 
               \Gamma_R \, \det \mbox{\boldmath $K$}_{N_CN_C}(0)
        \, \Bigr] 
\;.
\label{eq:determinant_Ginv}
\end{align}
Here $\mbox{\boldmath $K$}_{ij}(0)$ is 
a $(N_C-1) \times (N_C-1)$  derived 
from $\mbox{\boldmath $K$}(0)$ by 
deleting the $i$-th row and the $j$-th column.
Similarly, $\mbox{\boldmath $K$}_{11}^{N_CN_C}(0)$ is 
a $(N_C-2) \times (N_C-2)$ matrix obtained from $\mbox{\boldmath $K$}(0)$ 
by deleting the first and the $N_C$-th rows, 
and the first and the $N_C$-th columns. 
At $T=0$, the dc conductance is determined by the Green's function 
which connects the two leads,  
$g_{N_C}^{\phantom{\dagger}} = \left(2 e^2 / h\right)  
    4\Gamma_R \Gamma_L \left| G_{N_C 1}^{+}(0)\right|^2$.  
It can also be expressed in terms of the scattering matrix
\begin{equation}
g_{N_C}^{\phantom{\dagger}} \,=\, {2 e^2 \over h} \   
          4\, \Gamma_L \Gamma_R \,
        {
         \,\bigl[\,
           \det \mbox{\boldmath $K$}_{1N_C}(0) 
        \,\bigr]^2 
        \over
\left| \det \left\{\mbox{\boldmath $G$}^+(0) \right\}^{-1} \right|^2}
         \;. 
\label{eq:cond} 
\end{equation}
Furthermore  at $T=0$, 
 the charge displacement can be determined by the Friedel sum rule, 
\begin{equation}
\Delta N_{\rm tot} \,=\,
- {2 \over \pi }\ \mbox{Im}\,  
\log \left[\, 
{ \det \left\{\mbox{\boldmath $G$}^+(0)\right\}^{-1}
}\,\right]  \;.
\label{eq:Friedel}
\end{equation}
Particularly for the constant density of states, 
eq.\ (\ref{eq:Friedel}) corresponds to the charge displacement defined by   
$\Delta N_{\rm tot} = \sum_{i=1}^{N_C}  \langle n_{d,i}\rangle$.\cite{tanaka1}
Recently, a related formulation which takes into account 
the self-energy corrections using the effective Hamiltonian  
has also been applied to a finite ring with a magnetic flux.\cite{RejecRamsak}

\subsection{Conductance at half-filling}
\label{subsec:half-filling}

Specifically at half-filling $\epsilon_d = -U/2$, 
the matrix  elements of 
 $\mbox{\boldmath $K$}(0) = \{ \widetilde{t}_{ij}\, \}$ become zero 
for $i$ and $j$ belonging to the same sublattice, i.e., 
$|i-j| =0,\, 2,\, 4,\, \ldots$.
Thus in this case, $\mbox{\boldmath $K$}(0)$ has a checkered structure, 
and it causes the even-odd dependence on the number of 
the interacting sites $N_C$.

For even $N_C$ ($=2M$), 
eq.\ (\ref {eq:cond}) can be rewritten in the form
\begin{align}
 g_{2M}^{\phantom{\dagger}} &\,=\,  {2 e^2 \over h} \ 
           { \Gamma_L\, \Gamma_R \  \widetilde{v}_{C}^{\,2} 
             \over
\left[\, \left(\Gamma_L\, \Gamma_R + \widetilde{v}_{C}^{\,2}
              \right)/2 \,\right]^2} \;, 
\label{eq:g_2_eff}
\\
\widetilde{v}_{C}^{\,2} &\,=\, 
- \,{\det \mbox{\boldmath $K$}(0) \over
   \det \mbox{\boldmath $K$}_{11}^{N_CN_C}(0)} 
\;,
\label{eq:v_eff}
\end{align}
where we have used the relations which can be deduced from the 
checkered structure of $\mbox{\boldmath $K$}(0)$:  
$\det \mbox{\boldmath $K$}_{N_CN_C}(0) =0$,   
          $\det \mbox{\boldmath $K$}_{11}(0) = 0 $, 
 and $ \left[\,\det \mbox{\boldmath $K$}_{N_C1}(0)\,\right]^2 =  
-\det \mbox{\boldmath $K$}(0)
\, \det \mbox{\boldmath $K$}_{11}^{N_CN_C}(0)$.  
Specifically, for free electrons at $U=0$,  
$\mbox{\boldmath $K$}(0)$ is given simply 
by $-\mbox{\boldmath $H$}_C^0$, and 
 eq.\ (\ref{eq:v_eff}) yields $\widetilde{v}_{C}^{\,2}=t^{2}$.

For odd $N_C$ ($=2M+1$), 
the dc conductance, eq.\ (\ref{eq:cond}), can be expressed as
\begin{equation}
 g_{2M+1}^{\phantom{\dagger}} \,= \,  {2 e^2 \over h} \ 
           { \widetilde{\Gamma}_L \,\widetilde{\Gamma}_R   
             \over
            [\, (\widetilde{\Gamma}_L + \widetilde{\Gamma}_R)/2 \,]^2} \;, 
\label{eq:g_1_eff}
\end{equation}
where $\widetilde{\Gamma}_L = \lambda\,\Gamma_L$,    
$\widetilde{\Gamma}_R  = \Gamma_R \,/\, \lambda$, and 
\begin{equation}
 \lambda \,=\,   \sqrt{ 
                      \det \mbox{\boldmath $K$}_{11}(0) \over
                      \det \mbox{\boldmath $K$}_{N_CN_C}(0) 
                    } \;.
\label{eq:lambda}
\end{equation}
Here we have used the properties, 
   $\det \mbox{\boldmath $K$}(0) =0$,  
   $\det \mbox{\boldmath $K$}_{11}^{N_CN_C}(0)  =  0$,
   and  
$ \left[\, \det \mbox{\boldmath $K$}_{N_C1}(0)\,\right]^2 =  
\det \mbox{\boldmath $K$}_{11}(0)\, \det \mbox{\boldmath $K$}_{N_CN_C}(0) $.
Especially, if the system has 
the inversion symmetry $\Gamma_L=\Gamma_R$ in 
addition to the electron-hole symmetry,  
the parameter defined in eq.\ (\ref{eq:lambda})  becomes $\lambda=1$, 
and then 
the perfect transmission occurs, $g_{2M+1}^{\phantom{\dagger}}={2 e^2 / h}$,    for any value of $M$ and  $U$.

\begin{figure}[tb]
\begin{center}

%% N_C=3 Hubbard (NRG)
%% \hspace{-4cm}
\setlength{\unitlength}{0.7mm}
\begin{picture}(130,40)(-12,0)
\thicklines

\put(15.5,23.5){\makebox(0,0)[bl]{\Large$t$}}
\put(15.5,7){\makebox(0,0)[bl]{\Large$t$}}

%% \put(28,11){\makebox(0,0)[bl]{\Large$\bar{v}_L^{\phantom{0}}$}}
%% \put(27.5,26){\makebox(0,0)[bl]{\Large$\bar{v}_R^{\phantom{0}}$}}

\put(38,29){\makebox(0,0)[bl]{\Large$0$}}
\put(53,29){\makebox(0,0)[bl]{\Large$1$}}
\put(65,29){\makebox(0,0)[bl]{\Large$\cdots$}}
\put(82,29){\makebox(0,0)[bl]{\Large$N$}}

\put(-7,30){\makebox(0,0)[bl]{\Large \bf (a)}}
%% \put(5,30){\makebox(0,0)[bl]{\Large \bf (a)}}

%% \put(90,1){\makebox(0,0)[bl]{Left lead}}
%% \put(86,28.5){\makebox(0,0)[bl]{Right lead}}

\put(9,17.5){\line(2,1){14}}
\put(9,17.5){\line(2,-1){14}}
\put(10,17.5){\circle*{4}}

%%%%% \put(10,10){\line(0,1){15}}

%% \put(12,10){\line(1,0){11}}
%% \put(27,10){\line(1,0){11}}
\multiput(27,10)(3,0){4}{\line(1,0){1.5}}
\put(42,10){\line(1,0){11}}
\put(57,10){\line(1,0){11}}
\put(72,10){\line(1,0){11}}
\put(87,10){\line(1,0){6}}

%% \put(12,25){\line(1,0){11}}
%% \put(27,25){\line(1,0){11}}
\multiput(27,25)(3,0){4}{\line(1,0){1.5}}
\put(42,25){\line(1,0){11}}
\put(57,25){\line(1,0){11}}
\put(72,25){\line(1,0){11}}
\put(87,25){\line(1,0){6}}

%% \put(10,10){\circle*{4}} 
\put(25,10){\circle*{4}} 
\put(40,10){\circle{4}} 
\put(55,10){\circle{4}} 
\put(70,10){\circle{4}} 
\put(85,10){\circle{4}} 
%% \put(100,10){\circle{4}} 

%% \put(10,25){\circle*{4}} 
\put(25,25){\circle*{4}} 
\put(40,25){\circle{4}} 
\put(55,25){\circle{4}} 
\put(70,25){\circle{4}} 
\put(85,25){\circle{4}} 
%% \put(100,25){\circle{4}} 
\end{picture}

%% \rule{0cm}{2cm}

%% N_C=4 Hubbard (NRG)
\rule{0.1cm}{0cm}
%% \hspace{-4cm}
\setlength{\unitlength}{0.75mm}
\begin{picture}(130,40)(-12,0)
\thicklines

%% \put(90,1){\makebox(0,0)[bl]{Left lead}}
%% \put(86,28.5){\makebox(0,0)[bl]{Right lead}}

\put(16.5,27){\makebox(0,0)[bl]{\Large$t$}}
\put(16.5,12){\makebox(0,0)[bl]{\Large$t$}}
\put(5,16){\makebox(0,0)[bl]{\Large$t$}}

%% \put(28,11){\makebox(0,0)[bl]{\Large$\bar{v}_L^{\phantom{0}}$}}
%% \put(27.5,26){\makebox(0,0)[bl]{\Large$\bar{v}_R^{\phantom{0}}$}}

\put(38,29){\makebox(0,0)[bl]{\Large$0$}}
\put(53,29){\makebox(0,0)[bl]{\Large$1$}}
\put(65,29){\makebox(0,0)[bl]{\Large$\cdots$}}
\put(82,29){\makebox(0,0)[bl]{\Large$N$}}

%% \put(5,33){\makebox(0,0)[bl]{\Large \bf (b)}}
\put(-7,33){\makebox(0,0)[bl]{\Large \bf (b)}}

\put(10,10){\line(0,1){15}}

\put(12,10){\line(1,0){11}}
%% \put(27,10){\line(1,0){11}}
\multiput(27,10)(3,0){4}{\line(1,0){1.5}}
\put(42,10){\line(1,0){11}}
\put(57,10){\line(1,0){11}}
\put(72,10){\line(1,0){11}}
\put(87,10){\line(1,0){6}}

\put(12,25){\line(1,0){11}}
%% \put(27,25){\line(1,0){11}}
\multiput(27,25)(3,0){4}{\line(1,0){1.5}}
\put(42,25){\line(1,0){11}}
\put(57,25){\line(1,0){11}}
\put(72,25){\line(1,0){11}}
\put(87,25){\line(1,0){6}}

\put(10,10){\circle*{4}} 
\put(25,10){\circle*{4}} 
\put(40,10){\circle{4}} 
\put(55,10){\circle{4}} 
\put(70,10){\circle{4}} 
\put(85,10){\circle{4}} 
%% \put(100,10){\circle{4}} 

\put(10,25){\circle*{4}} 
\put(25,25){\circle*{4}} 
\put(40,25){\circle{4}} 
\put(55,25){\circle{4}} 
\put(70,25){\circle{4}} 
\put(85,25){\circle{4}} 
%% \put(100,25){\circle{4}} 

\end{picture}
\caption{Schematic pictures of discretized Hamiltonian for the NRG approach
for (a) $N_C =3$ and (b) $N_C = 4$}
\label{fig:model_NRG}
\end{center}

\end{figure}
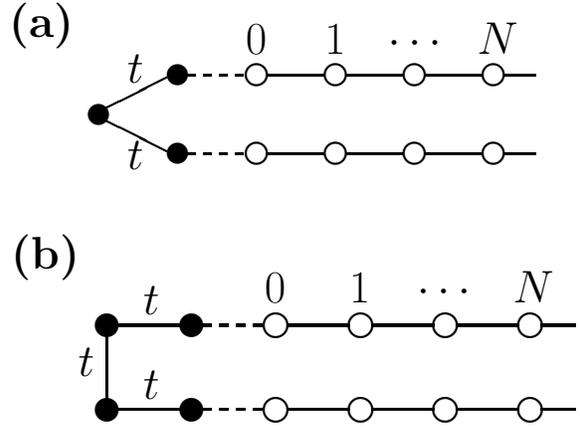

\section{Fixed-point Hamiltonian and conductance}
\label{sec:flow_vs_g}

In the NRG method, 
the conduction band can be modelled by a linear chain 
as shown in Fig.\ \ref{fig:model_NRG},
via a standard procedure of logarithmic discretization.
\cite{Wilson,KWW}
Then, to capture the low-energy behavior correctly,
 we use a sequence of the Hamiltonian $H_N$ defined by 
\begin{align}
H_N & \,=\,  \Lambda^{(N-1)/2}  
\left( \,
      {\cal H}_{d}  + {\cal H}_{U} 
 +  H_{\rm mix}^{\phantom{0}}  +   H_{\rm lead}^{(N)}
 \,\right) ,
\label{eq:H_N} 
\\
H_{\rm mix}^{\phantom{0}} & \,= \     \bar{v}_L^{\phantom{0}} \, 
       \sum_{\sigma}
\left(\,
f^{\dagger}_{0,L\sigma} d^{\phantom{\dagger}}_{ 1,\sigma}
\,+\, 
d^{\dagger}_{ 1,\sigma}  f^{\phantom{\dagger}}_{0,L\sigma} 
     \right)
     \nonumber \\
 & \quad  +   
        \bar{v}_R^{\phantom{0}} \, 
       \sum_{\sigma} 
       \left(\,
f^{\dagger}_{0,R \sigma} d^{\phantom{\dagger}}_{ N_C, \sigma} 
   \,+\, 
d^{\dagger}_{ N_C, \sigma} f^{\phantom{\dagger}}_{0,R \sigma} 
         \,  \right)  \;,
\label{eq:H_mix_NRG}
\\
H_{\rm lead}^{(N)} &\,=\,
D\,{1+1/\Lambda \over 2} \,
\sum_{\nu=L,R}
\sum_{\sigma}
\sum_{n=0}^{N-1} 
\, \xi_n\, \Lambda^{-n/2}
\nonumber \\
& \qquad \quad \times  
\left(\,
  f^{\dagger}_{n+1,\nu\sigma}\,f^{\phantom{\dagger}}_{n,\nu\sigma}
  +  
 f^{\dagger}_{n,\nu\sigma}\, f^{\phantom{\dagger}}_{n+1,\nu\sigma}
 \,\right) \;,
\label{eq:H_lead_NRG}
\end{align}
where $D$ is the half-width of the conduction band.
The hopping matrix elements   
  $\bar{v}_{\nu}^{\phantom{0}}$ and 
 $\xi_n$  are defined by
\begin{align}
  \bar{v}_{\nu}^{\phantom{0}}
&\,=\, \sqrt{ \frac{2D\,\Gamma_{\nu} A_{\Lambda}}{\pi} }
\;,
\qquad
A_{\Lambda}  \,=\,  \frac{1}{2}\, 
 {1+1/\Lambda \over 1-1/\Lambda }
\,\log \Lambda
\;, 
\label{eq:A_lambda}
\\
\xi_n &\,=\,    
{ 1-1/\Lambda^{n+1}  
\over  \sqrt{1-1/\Lambda^{2n+1}}  \sqrt{1-1/\Lambda^{2n+3}} 
} 
\;.
\label{eq:xi_n}
\end{align}
The factor $A_{\Lambda}$ is needed to compare 
the discretized model with the original Hamiltonian 
eq.\ (\ref{eq:H}) precisely, and 
it behaves as $A_{\Lambda}\to 1$ in the continuum limit $\Lambda \to 1$.
\cite{KWW,SakaiShimizuKasuya}
The low-lying energy states of the original Hamiltonian ${\cal H}$ 
can be deduced from those of $\Lambda^{-(N-1)/2} H_N$ for large $N$.

In the following we concentrate 
on the case $\Gamma_L = \Gamma_R$ ($\equiv \Gamma$), 
 where the couplings to the two leads 
 are symmetric $\bar{v}_L^{\phantom{0}} = \bar{v}_R^{\phantom{0}}$ 
 ($\equiv \bar{v}$).
In the discretized Hamiltonian $H_N$ in eq.\ (\ref{eq:H_N}), 
the matrix elements 
$t$ and $\bar{v}$ has multiplied by $\Lambda^{(N-1)/2}$.
As shown in the next section, 
for large $N$ the low-lying energy states of 
the many-body Hamiltonian $H_N$ converge 
to the states which have one-to-one correspondence to 
the quasi-particles of a local Fermi liquid.
It enables us to deduce the matrix elements of 
$\mbox{\boldmath $H$}_C^{\rm eff}$, which are defined in Eq.\ (\ref{eq:K}), 
 from the NRG spectrum.

At the fixed point,
the low-energy spectrum of the many-body Hamiltonian 
$H_N$ can be reproduced by the one-particle Hamiltonian consisting 
of $\mbox{\boldmath $H$}_C^{\rm eff}$ 
and the two finite leads;  
\begin{equation}
H_{\rm qp}^{(N)}  \, = \, 
\Lambda^{(N-1)/2}  
\left( \,
H_C^{\rm eff} + H_{\rm mix} + H_{\rm lead}^{(N)} 
\,\right) 
\;,
\label{eq:H_qp}
\end{equation}
where $H_C^{\rm eff} =  - \sum_{ij=1}^{N_C} 
\widetilde{t}_{ij} \, d^{\dagger}_{i \sigma} d^{\phantom{\dagger}}_{j \sigma}$.
It describes the free quasi-particles in the cluster 
with $N_C + 2(N+1)$ sites,  
and the corresponding  Green's function can be written as
\begin{align}
& \left\{\mbox{\boldmath $G$}_{\rm qp}(\omega)\right\}^{-1} 
\,\equiv \,
\nonumber \\
& 
\Lambda^{(N-1)/2} \,  
 \Bigl[ \,
 \omega  \Lambda^{-(N-1)/2} \,\mbox{\boldmath $1$} -
 \mbox{\boldmath $H$}_C^{\rm eff}  
 -  \Lambda^{(N-1)/2} \, \mbox{\boldmath $V$}_{\rm mix}^+(\omega) 
\, \Bigr ] 
\;. 
\end{align}
Here we have not included the renormalization 
factor $\partial \mbox{\boldmath $\Sigma$}/\partial \omega$,
because at $T=0$ it does not affect the dc conductance 
and charge displacement defined in eqs.\ (\ref{eq:cond}) 
and  (\ref{eq:Friedel}).
An eigenvalue $\varepsilon^*$ of $H_{\rm qp}^{(N)}$ satisfies the equation  
$\det \left\{\mbox{\boldmath $G$}_{\rm qp}(\varepsilon^*)\right\}^{-1} = 0$, 
which can be written in a 
form similar to eq.\ (\ref{eq:determinant_Ginv}),  
\begin{align}
& \det \mbox{\boldmath $K$}_{11}^{N_CN_C}
      (\omega_N) 
      \bigl[\, \bar{v}^{2}  \Lambda^{(N-1)/2} 
      \mbox{\sl g}_N (\varepsilon^*) \,\bigr]^2
+ \, \det \mbox{\boldmath $K$}(\omega_N)
     \nonumber \\
   & 
- \bigl[ \det \mbox{\boldmath $K$}_{11}(\omega_N) 
         +  \det \mbox{\boldmath $K$}_{N_CN_C}(\omega_N) 
          \bigr] 
     \bigl[ \bar{v}^{2} 
    \Lambda^{(N-1)/2}\mbox{\sl g}_N (\varepsilon^*) \bigr]
          \nonumber \\ 
&
\, = \,  0 \;,
\label{eq:NRG_eigenvalue}
\end{align}
where $\omega_N  \equiv \varepsilon^*\,\Lambda^{-(N-1)/2}$.
The Green's function $\mbox{\sl g}_N (\omega)$ is 
introduced for an isolated lead with $N+1$ sites, 
and is defined with respect to the interface $n=0$. 
It can be expressed as $\mbox{\sl g}_N (\omega) = 
\sum_{m=0}^{N} |\varphi_m(0)|^2/(\omega- \epsilon_m)$,
where $\epsilon_m$ and $\varphi_m(n)$ 
are the eigenvalue and eigenstate for the isolated lead.
In the electron-hole symmetric case, 
Eq.\ (\ref{eq:NRG_eigenvalue}) can be simplified by using 
the properties described in Sec.\ \ref{subsec:half-filling}, 
as follows.

For even $N_C$ ($=2M$) and in the limit of large $N$, 
Eq.\ (\ref{eq:NRG_eigenvalue}) yields
\begin{equation}
      \Bigl[\, \bar{v}^{2}  \lim_{N\to\infty} \Lambda^{(N-1)/2} 
      \mbox{\sl g}_N (\varepsilon^*) \,\Bigr]^2
\, = \, 
 - \, {\det \mbox{\boldmath $K$}(0)
     \over
     \det \mbox{\boldmath $K$}_{11}^{N_CN_C}(0)} \;. 
\label{eq:NRG_eigenvalue_even}
\end{equation}
Thus, the parameter $\widetilde{v}_{C}^{\,2}$ defined in
 eq.\ (\ref{eq:v_eff}) can be related 
to the eigenvalue $\varepsilon^*$, as 
\begin{equation}
{
\widetilde{v}_{C}^{\,2}
\over
\Gamma^2} \,=\,
 \left({\bar{v}^2 \over \Gamma D } \right)^2 
 \Bigl[\,\lim_{N\to\infty} D \Lambda^{(N-1)/2} \mbox{\sl g}_N (\varepsilon^*) 
 \,\Bigr]^2 
  \;.
\label{eq:FL_parameter}
\end{equation}
The prefactor in the right-hand side
can also be written as  $\bar{v}^2/(\Gamma D)  =  2 A_{\Lambda} / \pi$ 
by using eq.\ (\ref{eq:A_lambda}).
Note the function, 
$\lim_{N\to\infty} \Lambda^{(N-1)/2} \mbox{\sl g}_N (\omega)$,   
depends on whether $N$ is even or odd.
For even $N_C$, the fixed-point 
eigenvalue $\varepsilon^*$ depends on $U/t$ and $\Gamma/t$,
and the fixed-point Hamiltonian can be written in  the form   
\begin{equation}
H_{\rm qp}^{(N)}  
\, = \, 
\sum_{\sigma}\sum_{l=1}^{N_{\rm qp}} \varepsilon_l^* 
\left( 
\alpha_{l\sigma}^{\dagger} \alpha_{l\sigma}^{\phantom{\dagger}}
- 
\beta_{l\sigma}^{\dagger} \beta_{l\sigma}^{\phantom{\dagger}}
\right) 
\;,
\label{eq:Hqp_even}
\end{equation}
where $N_{\rm qp} = N_C/2 + N +1$.
One can determine $\widetilde{v}_{C}^{\,2}/\Gamma^2$
by substituting the value of $\varepsilon_l^*$ deduced from the 
NRG results into eq.\ (\ref{eq:FL_parameter}),
and then the dc conductance can be obtained from eq.\ (\ref{eq:g_2_eff}). 
Our formulation to deduce the conductance 
is analogous to the method used for the asymmetric Anderson impurity
to determine the local charge from 
 the fixed-point eigenvalue and Friedel sum rule.\cite{KWW2}

The situation is quite different for odd $N_C$  ($=2M+1$). 
In this case, eq.\ (\ref{eq:NRG_eigenvalue}) yields two separate branches 
of low-energy states for large $N$,
\begin{equation}
\lim_{N\to\infty} \Lambda^{(N-1)/2} \mbox{\sl g}_N (\varepsilon^*) 
\, = \, 0 \;, 
\label{eq:1_N}
\end{equation}
and
\begin{equation}
\Bigl[\, 
\lim_{N\to\infty} \Lambda^{(N-1)/2} \mbox{\sl g}_N (\varepsilon^*) 
\,\Bigr]^{-1}  \, = \, 0 \;.
\label{eq:0_N}
\end{equation}
These can be deduced from 
the behavior of the determinants for small $\omega_N$, 
$\det \mbox{\boldmath $K$}_{11}^{N_CN_C}(\omega_N) \propto \omega_N$  
   and  $\det \mbox{\boldmath $K$}(\omega_N) \propto \omega_N$.  
These two branches imply that 
the eigenvalue $\varepsilon^*$ does not depend on $U$, $t$ and $\Gamma$. 
Specifically, eq.\ (\ref{eq:1_N}) corresponds to an isolated lead 
consisting of $N$ sites starting from $n=1$ and ending at $n=N$,
while eq.\ (\ref{eq:0_N}) corresponds to
another lead with size $N+1$ that includes the site $n=0$.
One way to interpret the fixed point is along the lines of  
 the original work of Wilson\cite{Wilson}, as a strong coupling 
 fixed point, such that 
a single site at $n=0$ is removed from one of the leads 
to join the interacting sites and 
to form a singlet ground state 
for this cluster consisting of $N_C+1$ sites.
However, here we derived  eqs.\ (\ref{eq:1_N}) and (\ref{eq:0_N})
 for the {\em connected} chain of $N_C + 2(N+1)$ sites 
and have interpreted the fixed-point using the Hamiltonian 
 $H_{\rm qp}^{(N)}$, which is defined in eq.\ (\ref{eq:H_qp}) and
 can be diagonalized as
\begin{equation}
H_{\rm qp}^{(N)}  
\, = \, 
\sum_{\sigma} \varepsilon_0^* \,
\alpha_{0\sigma}^{\dagger} \alpha_{0\sigma}^{\phantom{\dagger}}
\,+\,
\sum_{\sigma}\sum_{l=1}^{N_{\rm qp}'} \varepsilon_l^* 
\left( 
\alpha_{l\sigma}^{\dagger} \alpha_{l\sigma}^{\phantom{\dagger}}
- 
\beta_{l\sigma}^{\dagger} \beta_{l\sigma}^{\phantom{\dagger}}
\right) 
\;,
\label{eq:Hqp_odd}
\end{equation}
where $\varepsilon_0^*=0$  and $N_{\rm qp}' = (N_C-1)/2 + N +1$.
This Hamiltonian links directly to the Fermi-liquid behavior, 
as the quasi-particles defined in this connected chain
are in one-to-one correspondence with the single-particle
excitations  of the non-interacting system ($U=0$). This leads to
 more natural description for the ground state properties 
than the strong coupling interpretation which involves breaking the
chain by effectively removing two sites.\cite{HewsonOguriMayer}
As we show in the next section,
the low-lying energy states of the many-body Hamiltonian $H_N$ 
reproduce the energy spectrum determined by   
eqs.\ (\ref{eq:1_N}) and (\ref{eq:0_N}).
The same behavior was seen in the single impurity case.\cite{Wilson,KWW}
Our numerical results confirm the Fermi-liquid behavior for $N_C=3$, 
and justify that the assumptions made in deducing 
the unitary-limit transport result, 
$g_{2M+1}^{\phantom{\dagger}} =  {2 e^2/ h}$, 
for the case  $N_C=3$.

\begin{fullfigure}[tb]
\begin{center}
\leavevmode
\includegraphics[ width=0.9\linewidth]{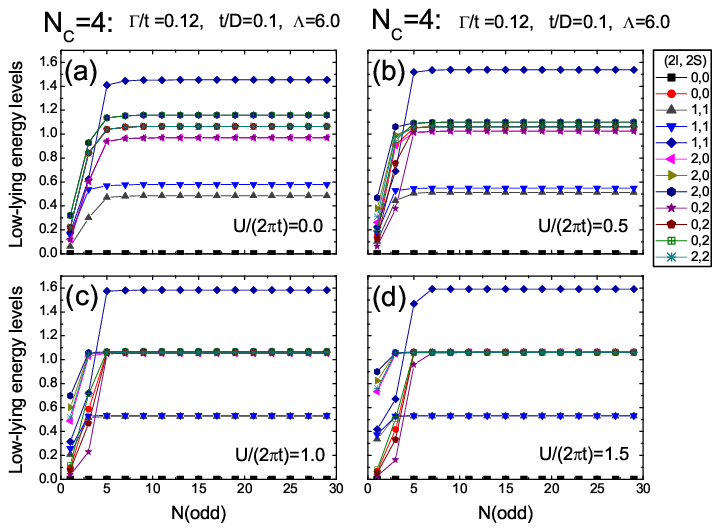}
\caption{
Low-lying energy levels of $H_N/D$ for $N_C=4$
as a function of odd $N$ (up to 29) for several values 
$U/(2\pi t)$: (a) $0.0$, (B) $0.5$, (c) $1.0$, and (d) $1.5$.
Here  $\epsilon_d = -U/2$,  $\Gamma/t = 0.12$, $t/D =0.1$  
and  $\Lambda=6.0$.    
The eigenvalues are measured from the ground-state energy for each $N$.
The label ($2I$, $2S$) corresponds to 
the total axial charge $I$ and spin $S$.
 For $N \gtrsim 7$, the levels approach to the fixed point values. 
}
\label{fig:energy_flow_4}
\end{center}
\end{fullfigure}

\section{NRG results}
\label{sec:NRG_results}

In the electron-hole symmetric case,
the Hamiltonian $H_N$ has a global $SU(2)$ symmetry 
of the axial charge,\cite{CoxZawadowski}    
which is specified by the generators
\begin{align}
\widehat{I}_z &\,= \,\sum_{i=1}^{N_C} 
\,\frac{1}{2} 
\left(\,
d_{i\uparrow}^{\dagger}
d_{i\uparrow}^{\phantom{\dagger}}
+
d_{i\downarrow}^{\dagger}
d_{i\downarrow}^{\phantom{\dagger}} -1
\,\right) 
\nonumber \\
& \ \  +   
\sum_{\nu=L,R}
\sum_{n=0}^{N}
\,\frac{1}{2}
\left(\,
f_{n,\nu\uparrow}^{\dagger}
f_{n,\nu\uparrow}^{\phantom{\dagger}}
+
f_{n,\nu\downarrow}^{\dagger}
f_{n,\nu\downarrow}^{\phantom{\dagger}} -1
\,\right) \;, 
\\
\widehat I_+ & \, = \,
\sum_{i=1}^{N_c} (-1)^i d_{i\uparrow}^{\dagger}
d_{i\downarrow}^{\dagger} +
\sum_{\nu=L,R}\sum_{n=0}^{N} (-1)^{\theta_{n\nu}}
f_{n,\nu\uparrow}^{\dagger}
f_{n,\nu\downarrow}^{\dagger} \;, \\
\widehat{I}_- & \, = \,
\sum_{i=1}^{N_c} (-1)^{i} 
d_{i\downarrow}^{\phantom{\dagger}}
d_{i\uparrow}^{\phantom{\dagger}} + 
\sum_{\nu=L,R}
\sum_{n=0}^{N} (-1)^{\theta_{n\nu}}
f_{n,\nu\downarrow}^{\phantom{\dagger}}
f_{n,\nu\uparrow}^{\phantom{\dagger}} \;.
\end{align}
Here $\theta_{n,L} \equiv -n$ for the left lead, 
and $\theta_{n,R} \equiv N_C + n +1$ for the right lead, 
so that  
the factor $(-1)^{\theta_{n \nu}}$ becomes 
$+1$ or $-1$ depending on whether the  
site labeled by ($n,\nu$) is in an even or odd sublattice.
The $z$ component of the axial charge 
corresponds to the total charge,  $\widehat{Q} = 2 \widehat{I}_z$.
Furthermore, 
the operators  $\widehat{I}_z$ and $\widehat{I}_{\pm}$  satisfy 
the commutation relations identical to those of 
the total spin operators $\widehat{S}_z$ and $\widehat{S}_{\pm}$.
Thus, using the symmetry 
of $SU(2)_{\rm spin} \times SU(2)_{\rm axial}$, 
the eigenstates can be classified according to 
the quantum numbers for the operators 
 $\widehat{S}_z$,
 $\widehat{S}^2$,  
 $\widehat {I}_z$, and 
 $\widehat{I}^2 \equiv
 \widehat{I}_z^2 
 + ( \widehat{I}_+ \widehat{I}_- + \widehat{I}_- \widehat{I}_+)/2 
$, as  
\begin{equation}
H_N\, |I,I_z,S,S_z\, ; r  \rangle_N 
\, = \,  E_{N,I,S,r}\, |I,I_z,S,S_z\, ; r  \rangle_N 
\;.
\end{equation}
The use of the $SU(2)_{\rm spin} \times SU(2)_{\rm axial}$ symmetry 
 has a great numerical advantage  [see also, Appendix].
One can save the eigenstates to be retained in the process 
of the NRG iteration.
Particularly in the multi-channel systems, 
such as the one we are 
considering (2-channel in our case),
the number of low-energy states to be retained   
increases exponentially with the number of the channels. 
Thus the reduction of the Hilbert space 
 helps to improve the numerical accuracy. 
After using the two $SU(2)$ symmetries, 
we have retained typically 1000 low-energy states 
in the NRG calculations for the Hubbard model with $N_C =3$ and $4$.
Under this condition  the truncation occurs first at $N=2$,
namely $H_N$ has been diagonalized exactly up to $N=1$ where 
the total number of the sites in the cluster   
is $7$ and $8$ for $N_C=3$ and $4$, 
respectively, as shown in Fig.\ \ref{fig:model_NRG}. 
Since two new sites at $n=N+1$ are included 
in each step of the recursive procedure,
the dimension of Hilbert space becomes typically $4^2$ 
larger than the number of the states retained in the previous step
apart from some reductions due to the symmetries. 
To overcome the influence of the truncation, 
which starts not so far from the interacting region, 
we have concentrated on the case in which $\Gamma/t$ is 
small ($\simeq 0.1$) and have used a rather large value for 
the discretization parameter $\Lambda =6.0$.   
The hopping matrix between the interacting sites has 
taken to be $t/D=0.1$ in the calculations.

  The use of an additional inversion symmetry 
   $\Gamma_L^{\phantom{0}} = \Gamma_R^{\phantom{0}}$ 
 does the change the total number of the basis states to be retained 
 in the interacting case $U \neq 0$,
 although each subspace labeled by $I$ and $S$ can be divided up 
 into two segments. 
 The inversion symmetry 
 can be employed by introducing the bonding and antibonding 
 orbits $a_{n,\pm,\sigma}^{\phantom{\dagger}} =
( 
f_{n,R,\sigma}^{\phantom{\dagger}}
\pm 
f_{n,L,\sigma}^{\phantom{\dagger}} 
)/\sqrt{2}$. However, for even $N_C$ 
these orbits make the axial charge nonlocal,
while the locality is preserved for odd $N_C$.  
We have also performed the calculations 
using these orbits for odd $N_C$. 
It makes the computer time  somewhat shorter, 
but is not essential for improving the numerical accuracy.

\subsection{Results for $N_C = 4$}

\begin{figure}[bt]
\begin{center}
\leavevmode
%% \includegraphics[width=0.9 \linewidth, clip, 
%% trim = 0cm 0.5cm 0cm 0.3cm]{e1_e2.eps}
\includegraphics[width=1.0 \linewidth, clip, 
trim = 0cm 0.5cm 0cm 0.3cm]{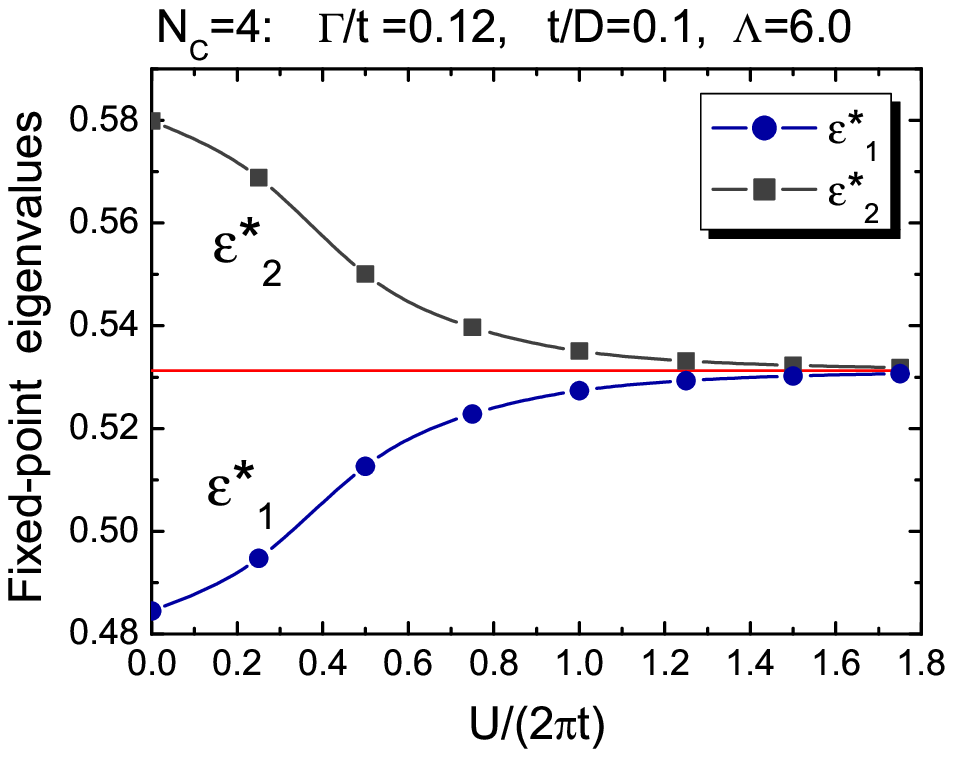}
\caption{ The $U$-dependence of $\varepsilon_1^*$ 
and $\varepsilon_2^*$ for the 4-site Hubbard model for odd $N$. 
For large $U$, these two approach to a single value $0.5312$, 
which corresponds to the lowest pole of the Green's function 
of the isolated lead,  
$
\lim_{N\to\infty} \Lambda^{(N-1)/2} \mbox{\sl g}_N (\omega) 
$, for odd $N$.
}
\label{fig:e1_e2}
\end{center}
\end{figure}

\begin{table}[b]
\begin{center}
\begin{tabular}{l}
\hline 
\hline 
$E_1^* \,=\, \varepsilon_1^*$ \\ 
$E_2^* \,=\, \varepsilon_2^*$ \\
$E_3^* \,=\, 2\, \varepsilon_1^*$ \\  
$E_4^* \,=\, \varepsilon_1^* \,+ \,\varepsilon_2^*$ \\  
$E_5^* \,=\, 2\, \varepsilon_2^*$ \\ 
$E_6^* \,=\, 3\, \varepsilon_1^*$ \\  
%% $E_7^* \,=\, 2 \,\varepsilon_1^* \,+\, \varepsilon_2^*$ \\  
%% $E_8^* \,=\, \varepsilon_1^* \,+\, 2 \varepsilon_2^*$ \\  
%% $E_9^* \,=\, 3\, \varepsilon_2^*$ \\  
%% $E_{10}^* \,=\, 4\, \varepsilon_1^* $  \\
\hline
\hline 
\end{tabular}
\end{center}
\caption{ Low-lying fixed-point eigenvalues 
of $H_N$ for $N_{\rm HUB} =4$, for odd $N$. 
%% Here $\Lambda=6$, $t/D=0.1$, $\Gamma/t=0.12$, and  $U/(2\pi t)$. 
}
\label{tab:N_C=4 for odd N}
\end{table}

\begin{figure}[tb]
\begin{center}
\leavevmode
%% \includegraphics[width=0.9 \linewidth, clip, 
%% trim = 0cm 0.5cm 0cm 0.3cm]{cond4svc.eps}
\includegraphics[width=1.02 \linewidth, clip, 
trim = 0.1cm 0.5cm 0cm 0.3cm]{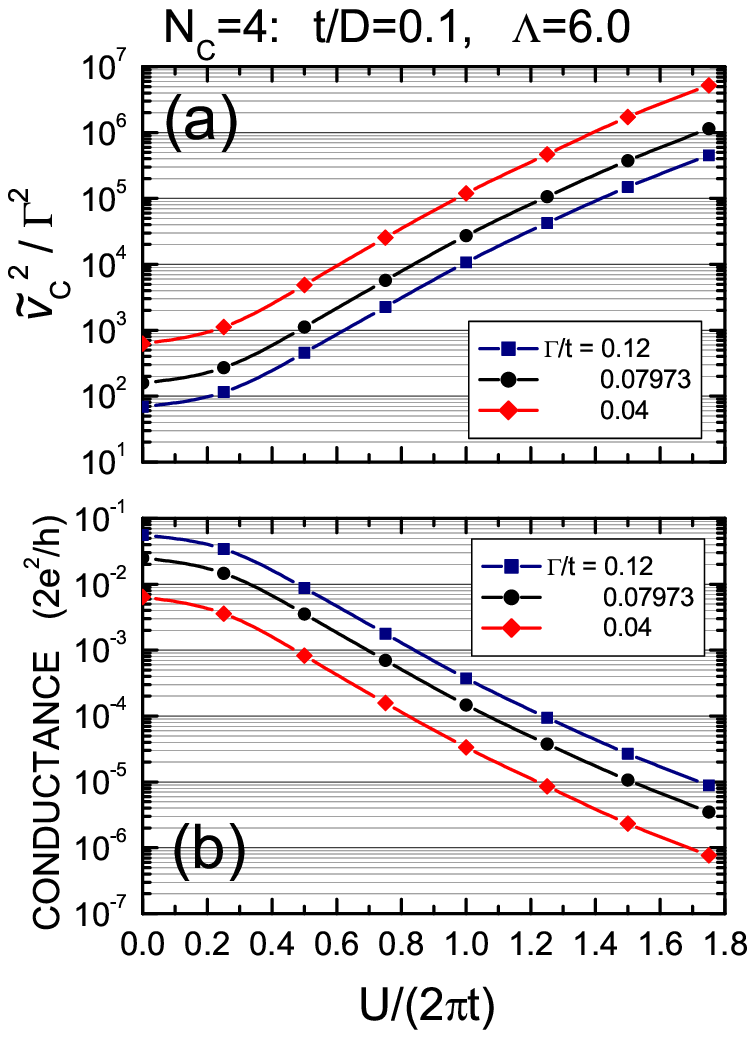}
\caption{
The $U$-dependence of (a) $\widetilde{v}_C^2$ 
  and (b) 
 the dc conductance $g_{N_C}^{\phantom{+}}$  
for the 4-site Hubbard model
for several values of the hybridization: 
$\Gamma/t=0.04$, $0.07973$, and $0.12$. 
The parameter $\widetilde{v}_C^2$  
is deduced form the fixed-point eigenvalue $\varepsilon_l^*$ 
by using eq.\ (\ref{eq:FL_parameter}). 
}
\label{fig:NRG4S_cond_vs}
\end{center}
\end{figure}

\begin{fullfigure}[tb]
\begin{center}
\leavevmode
\includegraphics[width=0.9\linewidth]{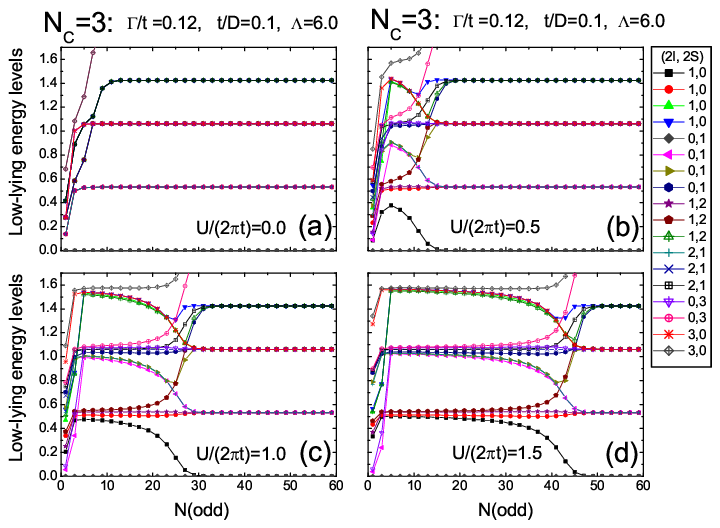}
\caption{
Low-lying energy levels of $H_N/D$ for $N_C=3$ 
as a function of odd $N$ (up to $59$) for several values 
$U/(2\pi t)$: (a) $0.0$,  (b) $0.5$, (c) $1.0$, and  (d) $1.5$.
Here  $\epsilon_d = -U/2$,  $\Gamma/t = 0.12$, $t/D =0.1$  
and  $\Lambda=6.0$.    
The eigenvalues are measured from the ground-state energy for each $N$.
The label ($2I$, $2S$) corresponds to 
the total axial charge $I$ and spin $S$.
The size of the lead, $N^*$, required to get the energy levels closed 
to the fixed-point values increases with $U$.
}
\label{fig:energy_flow_3}
\end{center}
\end{fullfigure}

We first of all consider the Hubbard chain of the size $N_C=4$,
which can be regarded as a simplest case for an even interacting chain.
In Fig.\ \ref{fig:energy_flow_4}, 
the low-lying  energy levels of $H_N$ 
for $\Gamma/t=0.12$ are plotted as a function of odd $N$, 
where $N+1$ is even, for several values of the Coulomb interaction: 
$U/(2 \pi t)= 0.0$, $0.5$, $1.0$, and $1.5$.
 The eigenvalues are measured from the ground-state energy for each $N$.  
The flow of the energy levels converges to the fixed-point values 
for $N\gtrsim 10$ in these examples.
In the case of even $N_C$,
the fixed-point eigenvalues depend on $U$ and $\Gamma$.
The numerical results for the many-body low-lying states can be compared with 
the quasi-particle eigenstates defined in eq.\ (\ref{eq:Hqp_even}). 
We found that the many-body eigenvalues $E^*_i$ of $H_N$ shown 
in Fig.\ \ref{fig:energy_flow_4} have 
 correspondence to the excited states 
described by the quasi-particles with the energy $\varepsilon^*_l$, 
as summarized in Table \ref{tab:N_C=4 for odd N}.
The first two excitation energies of $H_N$, i.e.,
 $E^*_1$ and $E^*_2$, determine the two {\em one-particle} 
 energies $\varepsilon^*_1$ and $\varepsilon^*_2$,
respectively. The many-body eigenvalues above these two,
 $E^*_3$,  $E^*_4$, $\ldots$, agree well with those calculated 
 from $\varepsilon^*_1$ and $\varepsilon^*_2$ with  
the assignments given in the table \ref{tab:N_C=4 for odd N}. 
We have confirmed that 
the two quantum numbers, the total axial charge $I$ and spin $S$, 
for the low-energy eigenstates are consistent with these assignments.
This feature of the low-lying energy states is 
similar to that for the Kondo and Anderson models,\cite{Wilson,KWW}
and means that the low-energy properties can be 
described by the local Fermi-liquid theory.
Specifically, for even $N_C$, the 
fixed-point Hamiltonian $H_{\rm qp}^{(N)}$ defined 
in eq.\ (\ref{eq:H_qp}) can be separated into 
a couple of the chains when the system has an inversion symmetry, 
and each of the chains can be mapped onto 
a noninteracting version of the asymmetric Anderson model.\cite{KWW2}
In Fig.\ \ref{fig:e1_e2}, the $U$ dependence of 
 $\varepsilon^*_1$ and $\varepsilon^*_2$ are shown 
 for the parameter set; 
  $\Gamma/t=0.12$, $t/D=0.1$, and $\Lambda=6$. 
The difference between  $\varepsilon^*_1$ and $\varepsilon^*_2$ 
 decreases with increasing $U$. 
 This tendency links with the behavior of  
 the other excitation energies shown in  Fig.\ \ref{fig:energy_flow_4},
 namely  $E^*_3$, $E^*_4$, and $E^*_5$ become close to each other with 
 increasing $U$. 
For large $U$,  both $\varepsilon^*_1$ and $\varepsilon^*_2$ approach
to $0.5312$ which corresponds to the smallest pole of 
$
\lim_{N\to\infty} \Lambda^{(N-1)/2} \mbox{\sl g}_N (\omega) 
$ for odd $N$.  It means that in the limit of $U \to \infty$ 
the low-energy states are determined 
by those of the isolated leads. 

Substituting the value of $\varepsilon^*_1$ or $\varepsilon^*_2$ into 
eq.\ (\ref{eq:FL_parameter}) and taking $N=99$ 
for the  Green's function of the isolated lead, we obtain 
the parameter $\widetilde{v}_C$ defined in eq.\ (\ref{eq:v_eff}), 
and then determine the dc conductance via eq.\ (\ref{eq:g_2_eff}).
The results are plotted against $U/(2 \pi t)$ in 
Fig.\ \ref{fig:NRG4S_cond_vs}  for several values of hybridization; 
$\Gamma/t=0.04$, $0.07973$, and $0.12$. 
The value of $\widetilde{v}_C$ determined from $\varepsilon^*_1$ 
agree with that determined from $\varepsilon^*_2$.
Furthermore, we have also calculated $\widetilde{v}_C$ using 
the fixed-point eigenvalues for even $N$ and the Green's function 
$\Lambda^{(N-1)/2} \mbox{\sl g}_N (\omega)$ for $N=100$.
The result again agrees with that deduced from the data for odd $N$. 
The parameter $\widetilde{v}_C$ increases with 
the Coulomb interaction $U$ reflecting the behavior of 
the fixed-point eigenvalue shown in Fig.\ \ref{fig:e1_e2}.
The results for the conductance 
are plotted in Fig.\ \ref{fig:NRG4S_cond_vs} (b). 
When the coupling with leads $\Gamma$ increases,
the conductance also increases in the parameter region we have examined. 
For large $U$, the conductance decreases exponentially with increasing $U$. 
This can be understood as a tendency towards 
the development of a Mott insulator gap. 
Namely, for large even $N_C$, the conductance is expected to 
show a behavior $g_{N_C}^{\phantom{\dagger}} 
\propto e^{-N_C/\xi}$, where $\xi \sim \hbar v_F / \Delta_{\rm gap}$ 
is a correlation length determined 
by the Hubbard gap $\Delta_{\rm gap}$ and Fermi velocity $v_F$.
This is because  $\Delta_{\rm gap} \propto U$ for large $U$, 
as can be seen in the Bethe ansatz results 
for one-dimensional Hubbard model.\cite{LiebWu}
Our previous results obtained with the 2nd order perturbation 
theory in $U$ are valid  qualitatively.\cite{ao9} 
However, the 2nd-order perturbation theory fails to reproduce 
the correct exponential dependence for large $U$,
and it has now been corrected with the non-perturbative NRG technique.

\subsection{Results for $N_C = 3$}

We next consider the Hubbard chain with the odd number of interacting sites.
In Fig.\ \ref{fig:energy_flow_3}, 
the flow of the low-lying eigenvalues of $H_N/D$ 
for $N_C=3$ is plotted as a function of odd $N$ for several
values of the Coulomb interaction 
$U/(2 \pi t)= 0.0$, $0.5$, $1.0$, and $1.5$.
The flow of the eigenvalues of $H_N/D$ is quite different 
from that for even $N_C$. 
In the case of odd $N_C$, a number of noninteracting sites in the leads 
are required to reach the fixed point 
that determines the low-energy properties. 
The number of NRG iterations $N^*$ that is needed to get 
the convergent results increases with $U$, 
and in the case of Fig.\ \ref{fig:energy_flow_3} 
it is estimated to be (b) $N^* \simeq 15$, (c) $N^* \simeq 30$, and 
(d) $N^* \simeq 50$. 
It means that 
there is a characteristic energy scale determined   
 by $T^* \simeq D\Lambda^{-(N^*-1)/2}$,
where the factor $\Lambda^{-(N^*-1)/2}$ is introduced to recover 
the original energy scale from  $H_N$ defined in eq.\ (\ref{eq:H_N}).
This characteristic energy scale is 
determined by width of the Kondo resonance, 
$T^* \sim T_K$, appearing at the Fermi level.

For $N \gtrsim N^*$, 
the fixed-point eigenvalues do not depend on 
$U$ and agree with those of the the noninteracting 
leads determined by eqs.\ (\ref{eq:1_N}) and (\ref{eq:0_N}). 
Namely, these low-lying many-body states, $E^*_i$, have 
the one-to-one correspondence with 
the quasi-particle states described by eq.\ (\ref{eq:Hqp_odd}).
The precise correspondence is summarized in Table \ref{tab:N_C=3},
and these assignments coincide with those
for the single Anderson impurity.\cite{KWW}
The first and third excited states, $E_1^*$ and $E^*_3$,
correspond to the first two {\em one-particle} states, 
the energy of which are given by 
$\varepsilon_1^*= 0.53124$ and $\varepsilon_2^*= 1.42546$. 
We have also confirmed for $N_C=3$ that
the fixed-point eigenvalues at $N \gtrsim N^*$
do not depend on whether $N$ is even or odd.
These feature show that the low-temperature properties 
at $T \lesssim T^*$  can described by the quasi-particles 
of the local Fermi theory, and it justifies the assumptions made 
in deducing the unitary-limit conductance, $g_{N_C}^{\phantom{0}}=2e^2/h$,  
at $T=0$ for odd $N_C$.

To capture the low-energy Kondo behavior at $T \lesssim T_K$ correctly, 
one needs to repeat the NRG iterations up to $N \gtrsim N^*$ as 
mentioned in the above. 
In other words, a sufficiently large number of the noninteracting sites 
are required for the reservoirs to make 
the finite-size energy separation smaller than $T^*$,
and a similar notice has recently been emphasized 
by several authors.\cite{Bonca,Cornaglia}
From the results obtained for a small cluster with size corresponding 
to $N \lesssim N^*$,
it is still possible to deduce the high-temperature properties 
at $T \gtrsim T_K$.\cite{Chiape,Busser}
In the NRG method the logarithmic discretization of 
the conduction band, eq.\ (\ref{eq:H_N}), 
yields the hopping matrix element that 
decreases exponentially with increasing $N$,\cite{Wilson,KWW,KWW2} 
and it makes the convergence to the fixed point efficient.

\begin{table}[bt]
\begin{center}
\begin{tabular}{l}
\hline 
\hline 
$E_1^* \,=\, \varepsilon_1^*$ \\ 
$E_2^* \,=\, 2\, \varepsilon_1^*$ \\  
$E_3^* \,=\, \varepsilon_2^*$ \\
$E_4^* \,=\, 3\, \varepsilon_1^*$ \\  
$E_5^* \,=\, \varepsilon_1^* \,+ \,\varepsilon_2^*$ \\  
$E_6^* \,=\, 4\, \varepsilon_1^*$ \\ 
%% $E_7^* \,=\, 2 \,\varepsilon_1^* \,+\, \varepsilon_2^*$ \\  
\hline
\hline 
\end{tabular}
\end{center}
\caption{ Low-lying fixed-point eigenvalues 
of $H_N$ for $N_{\rm HUB} =3$. 
%% Here $\Lambda=6$, $t/D=0.1$, $\Gamma/t=0.12$, and  $U/(2\pi t)$. 
}
\label{tab:N_C=3}
\end{table}

%%%%%%%%%%%%%%%%%%%%%%%%%%%%%%%%%%%%%%%%%%%%%%%%%%%%%%%%%%%%%%%%%%%%%%%

\section{Discussion and Summary}
\label{sec:remarks}

In the present work, 
we have clarified the difference in the transport properties 
between the Hubbard chain of even and odd $N_C$. 
Then,  what happens in the limit of large $N_C$?
It might sound somewhat puzzling 
since the ground state of the Hubbard chain 
is an insulating state in the thermodynamic limit. 
However, the existence of the energy scale $T^*\sim T_K$ for odd $N_C$ 
brings us the answer. 
For odd $N_C$  ($=2M+1$), there is the Kondo resonance at the Fermi level.
The width $T_K$ decreases when $N_C$ increases, 
and finally $T_K \to 0$ in the limit of $N_C \to \infty$ 
 because the Hubbard gap $\Delta_{\rm gap}$ evolving with increasing $N_C$
disturbs the electrons to screen the local moment. 
Thus, the value of the conductance $g_{N}^{\phantom{\dagger}}$  depends on 
the order of taking the limits of $N_C \to \infty$ and $T \to 0$.
For finite $N_C$, the Kondo behavior at  $T<T_K$ causes 
the unitary-limit behavior
\begin{equation}
 \lim_{M\to\infty} \left[\,\lim_{T\to 0} g_{2M+1}^{\phantom{\dagger}} 
 \,\right]\,=\, 2e^2/h \;.
\end{equation}
In contrast, when the thermodynamic limit $N_C\to \infty$ 
is taken first at small but finite $T$, the conductance must be determined by 
the Mott-Hubbard behavior at $T > T_K$,  as 
\begin{equation}
 \lim_{T\to 0} \left[\,\lim_{M\to\infty} g_{2M+1}^{\phantom{\dagger}}
  \,\right] \, = \, 0 \;.
\end{equation}
Therefore, 
in order to observe the unitary-limit behavior 
at an accessible temperature, the number of the interacting 
sites $N_C$ should not be so large. 
For even $N_C$ ($=2M$)  
there is no Kondo resonance at Fermi level, 
and the conductance does not depend on the limiting procedure 
\begin{equation}
\lim_{\scriptstyle M\to\infty \atop \scriptstyle T \to 0}
\, g_{2M}^{\phantom{\dagger}}
 \, = \, 0 \;.
\end{equation}
The two limits, $N_C \to \infty$ and $T \to 0$, considered here are 
analogous to the $k\to 0$ and $\omega \to 0$ limits 
of the vertex corrections for the interacting Fermi systems with 
the translational invariance.\cite{AGD}

In summary, we have studied the conductance through a finite Hubbard 
chain of the size $N_C$ connected to two noninteracting leads 
using the NRG method.
The results show that the low-lying energy states 
can be described by the quasi-particles of a local Fermi liquid.
We have also presented a formulation for deducing the dc conductance 
from the fixed-point Hamiltonian.
The results of the conductance for even $N_C$
show the expected exponential decay as a function of $U$ at half-filling.

\bigskip
\medskip

\section*{Acknowledgements}
One of us (AO) wishes to acknowledge 
the support by the Grant-in-Aid 
for Scientific Research from JSPS.
ACH wishes to thank the EPSRC(Grant GR/S18571/01) 
for financial support.
Numerical computation was partly performed 
at Yukawa Institute Computer Facility. 

\bigskip
\medskip

\appendix

\section{Reduced matrix element}
\label{sec:axial_charge}

In the electron-hole symmetric case,
the Wigner-Eckart theorem for the spin and axial charge 
$SU(2)_{\rm spin} \times SU(2)_{\rm axial}$ yields 
\begin{align}
& \!\!\!\!\!\!
 \langle 
 I,I_z,S,S_z\, ; r  
 | f_{n,\nu\sigma}^{\dagger} |
  I',I_z',S',S_z'\, ; r' 
 \rangle 
 \nonumber \\
&\,= \,
 \langle S', S_z' ; 1/2, \sigma | S, S_z \rangle \, 
 \langle I', I_z' ; 1/2, 1/2 | I, I_z \rangle
\nonumber \\
& \ \quad \times \, F_{n,\nu}(I,S;r|I',S';r' )  \;,
\end{align}
where the Clebsh-Gordan coefficient appears for the total spin
 $\langle S', S_z' ; 1/2, \sigma | S, S_z \rangle$ 
and for the total axial charge $\langle I', I_z' ; 1/2, 1/2 | I, I_z \rangle$.
The invariant matrix element $F_{n,\nu}(\alpha'|\alpha)$ has 
the following properties against the exchange of the arguments 
$\alpha'$ and $\alpha$; 
\begin{align}
& \!\!\!\!\!\!\!\!
F_{n,\nu}(I,S-1/2;r'|I-1/2,S;r ) 
\nonumber \\
&=\, (-1)^{n}
 \sqrt{2S+1 \over 2S}
\sqrt{2I \over 2I+1} \, 
\nonumber \\
& \quad \times 
 F_{n,\nu}(I-1/2,S;r|I,S-1/2;r') \,,
\nonumber \\
\\
& \!\!\!\!\!\!\!\!
F_{n,\nu}(I,S+1/2;r'|I-1/2,S;r ) 
\nonumber \\
&=\, (-1)^{n+1}
 \sqrt{2S+1 \over 2S+2}
 \sqrt{2I \over 2I+1} \, 
\nonumber \\
& \quad \times 
 F_{n,\nu}(I-1/2,S;r|I,S+1/2;r') \,.
\nonumber \\
\end{align}

\end{document}